\documentclass[aps,superscriptaddress,twocolumn,showpacs,preprintnumbers,amsmath,amssymb,showkeys,textcomp]{revtex4}
\usepackage[cp1251]{inputenc}
\usepackage{textcomp,amssymb,amsmath}
\usepackage{amstext,textcomp} 
\usepackage[dvips]{hyperref}
\usepackage[mathscr]{eucal}
\usepackage{longtable}
\usepackage{graphicx}
\usepackage[english]{babel}

\begin{document}
\title{Ferrimagnetic Spin Wave Resonance and Superconductivity in Carbon Nanotubes Incorporated in Diamond Matrix} 
\author{Dmitri Yerchuck (a), Yauhen Yerchak (b), Vyacheslav Stelmakh (b), Alla Dovlatova (c),  Andrey Alexandrov (c)\\
(a) - 	Heat-Mass Transfer Institute of National Academy of Sciences of RB,
Brovka Str.15, Minsk, 220072, dpy@tut.by \\(b) - Belarusian State University, Nezavisimosti Avenue 4, Minsk, 220030, RB\\
(c) - M.V.Lomonosov Moscow State University, Moscow, 119899}
\date{\today}
             
\begin{abstract} The phenomenon  of ferrimagnetic  spin wave resonance [uncompensated antiferromagnetic spin wave resonance] has been detected for the first time. It has been observed in carbon nanotubes, produced by high energy ion beam modification of diamond single crystals in $\langle{100}\rangle$ direction. Peculiarities  of   spin wave resonance observed allow to insist on the formation in given nanotubes of   $s^+$-superconductivity at room temperature, coexisting with uncompensated antiferromagnetic ordering.  
\end{abstract}

\pacs{71.10.-w, 73.63.Fg, 78.30.-j, 76.30.-v, 76.50.+g, 78.67.-n}
                             
\keywords{magnetism, nanotubes, superconductivity}
\maketitle
\section{Introduction and Background}
A number of theoretical and experimental  works  have been devoted to the studies of ferromagnetic and antiferromagnetic spin waves, including
resonance behavior. Theoretical  works are starting in 1930 from   pioneering work of Bloch \cite{Bloch}, where the ferromagnetic
spin waves were fundamentally studied. It has been found in particular, that ferromagnetic spin waves obey the $k^2$-dispersion law. At the same time, it was established, that antiferromagnetic spin waves have quite other - linear $k$-dispersion law. It was done by Hulthen. Hulthen has carried through the quantization of the
spin waves in an antiferromagnet and he has found for the first time $k$-dispersion law for antiferromagnetic spin waves \cite{Hulthen}  in 1936. The study of antiferromagnetic spin waves was extended   by Anderson \cite{Anderson} by inclusion of zero-point spin wave energy.   Anderson has  shown, that the exact ground-state energy eigenvalue of antiferromagnet is close to the energy of its approximate two-sublattice model. Anderson has  found, that the frequencies of the spin waves fall into two 
branches, and they
are
\begin{equation}
\label{eq1a}
\omega_k = d J S \sqrt{1 - {\gamma_k}^2},
\end{equation}

where $\gamma_k$ is
\begin{equation}
\label{eq2a}
\gamma_k = \sum\limits_{i = 1}^{d} \frac{\cos k_i}{d},
\end{equation}
$k_i$,  $i = \overline{1,d}$ are the components of the wave vector  $\vec{k}$, $d$ is dimensionality of the lattice, $S$ is spin value, $J$ is exchange coupling parameter. The equation (\ref{eq1a}) is reduced in the case of small amplitude values  of
vector  $\vec{k}$ to the following form
\begin{equation}
\label{eq3a}
\omega_k \sim d J S k,
\end{equation}
This dispersion law is coinciding with the law, obtained in \cite{Hulthen}, that is it really  quite different from the ferromagnetic
case. Anderson draws attention, that 
the differerence in dispersion laws is not the most
significant difference between the two types of spin
waves; it is the difference in amplitude per quantum of
excitation which leads to the more striking effects. It means, in particular, that it is required much more energy to excite an antiferromagnetic spin wave, than to excite a ferromagnetic spin wave.

Spin wave theory  was devoloped, for instance, in  the works  \cite{F_Keffer},  \cite{Keffer}, \cite{Ziman}, \cite{Nakamura}, \cite{Tani} and in  many others.
It has been shown in \cite{F_Keffer}, \cite{Keffer}, that the transformation to normal spin wave modes, used by Anderson for the case of the absence of external magnetic field is not suitable in the case of its presence. Independently, normal spin wave modes were derived using a different formalism by Ziman \cite{Ziman} and by Nakamura  \cite{Nakamura}.
A brief outline is given  in \cite{Keffer} for the spin-wave approximation to the near ground 
states of ferromagnetism and antiferromagnetism. Simple pictorial models of spin 
waves were introduced. These models clarify the striking difference between 
ferromagnetic spin waves, which obey the $k^2$-dispersion law, and 
antiferromagnetic spin waves, which obey the $k$-dispersion law.  
The frequency spectrum and the damping constant of the spin waves in the two-sublattice
antiferromagnetics were investigated in \cite{Tani} on the quantum-statistical basis by the use of the relaxation
function method. The interest for nanophysics represents the study of uniform spin wave modes in antiferromagnetic nanoparticles with uncompensated 
moments, performed   in the work by Bahl et al  \cite{Bahl}. According to \cite{Bahl}, in magnetic nanoparticles the uniform precession (q = 0 spin wave) mode gives 
the predominant contribution to the magnetic excitations.  The authors  have calculated the 
energy of the uniform mode in antiferromagnetic nanoparticles with uncompensated 
magnetic moments, using the coherent potential approximation. They have shown, that in the simple 
uniaxial case, the uncompensated moment has a profound effect on the excitation 
energy, but in the planar case it is much less significant. In fact it has been shown, that spin wave modes in antiferromagnetics and ferrimagnetics are obeing qualitatively to the same laws, the difference is the only quantitative. 

Especially interesting seem to be the resonance phenomena on 
spin waves. The  phenomenon of ferromagnetic spin wave resonance (FMSWR) was theoretically predicted by Kittel \cite{C_Kittel} and it was experimentally confirmed by Seavey   and Tannenwald \cite{Seavey_Tannenwald}. On the observation of antiferromagnetic  spin wave resonance was claimed   in \cite{Lui}, at that the authors have insisted, that their work is the first work in given field. The experiments were performed on epitaxial films of $MnF_2$. Let us remark, that the interpetation of rather interesting experimental  results, proposed in \cite{Lui}, seems to be incorrect.  The authors describe antiferromagnetic  spin wave resonance in the frame of the $k^2$-dispersion law, that is, in full contradiction with general spin wave theory, briefly above rewieved. The analysis of their results is embarrassed by inaccurate representation of experimental data in \cite{Lui}. The authors insist, that several SWR modes are unresolved, however concrete number of unresolved modes is not indicated. Taking into account the distance between the first and the second lines in low intensive line sequence in the spectrum of the the sample with the thickness in 0.98 ${\mu}m$, equaled to $\approx$ 50 G and the distance of the first line in given sequence from the position of main resonance mode  equaled to $\approx$ 83 G, then according to $k^2$-dispersion law authors' concept, the first line has to be the seveth mode. It is not in agreement with corresponding distance between the the seventh and the ninth modes in the spectrum of the sample with the thickness in 0.23 ${\mu}m$, since to the splitting  in $\approx$ 50 G in the spectrum of the sample with the thickness in 0.98 ${\mu}m$ has to correspond the splitting $\approx$ $(0.98/0.23)^2 \times 50$ G, that is 907.4 G. At the same time the evaluation of   analogous distance, that is the distance between  the seventh and the ninth modes from the spectrum of the sample with the thickness in 0.23 ${\mu}m$ gives the value in $\approx$ 575 G. We see, that the descrepance is large. Moreover, taking into account the linewidth of the modes, equaled to 8 G, the only the first and the third modes can be unresolved between themselves, but they must give the feature (shoulder or not very pronouced peak) on the wing of much more broad main resonance mode, all the more  the mode with number 5 has to be seen. To give the correct explanation for the spectra observed, the additional experimental data have to be represented  - the values of ratio of amplitudes of magnetic component of microwave field and the ratio of Q-factors, gain factors for intensity in the spectra of both the samples and also modulation amplitude and modulation frequency used. It can be, in particular, new quantum effect.  

Therefore, it is followed from above  given analysis, that  the work \cite{Lui} cannot be referred to the bibliography of the works, describing the phenomenon of AFMSWR.  At the same time the spin-wave spectrum in antiferromagnets was studied 
experimentally already in 60th years of the last century, at that the theory of AFMSWR was simultaneously developed. For instance, the role  of uniaxial tension on the spin-wave spectrum in easy-plane antiferromagnets was studied, \cite{Turov_E}, \cite{Borovik-Romanov}, \cite{Turov}. It was shown that
the effect of uniaxial tension in the basal plane of the antiferromagnet
crystal can be described by an effective magnetic field, at that
 the additional gap arising in the spin-wave spectrum has to be taken into consideration. The effect of uniaxial tension is rather strong. It has been shown, that even weak distortions can significantly modify the spin wave spectrum. 

Very interesting results on  AFMSWR are reported in \cite{Nidda}.  Spin wave resonance lines with extremely large wave numbers corresponding to wave vectors
$k$ in the range  $10^5 - 10^6 cm^{-1}$ were observed in thin plates of $FeBO_3$.
Spin-wave resonance was clearly observed in the temperature interval 30-250 K
 It was the low frequency branch of the
spin-wave spectrum, which was analysed at static magnetic field $H$, directed  transversely to $C_3$ crystal axis by the
 relation, taking  additionally  into account, in distinction from relation (\ref {eq1a}), Dzyaloshinsky field and magnetoelastic coupling

\begin{equation}\label{eq4a}
\omega_{1,k} = \gamma \sqrt{H (H + H_D) + H^2_{\Delta} + \alpha^2_1 k^2_1 + \alpha^2_2 k^2_2},
\end{equation}
where $\gamma$
 is the gyromagnetic ratio,  $H_D $ is the Dzyaloshinsky field, $H_{\Delta}$ is a parameter determined by magnetoelastic coupling,  $\alpha_1$ and $\alpha_2$  are non-uniform exchange
constants in the basal plane and along
the $C_3$-axis respecively, $k_1$ and $k_2$ are wave-vector components analogously in the basal plane and along
the $C_3$-axis respecively, external field $H$ is applied in the basal
plane of the crystal.

 We have to remark, that there is  optical analogue of AFMSWR, that is 
antiferroelectric spin wave resonance (AFESWR), which was theoretically described and experinmentally discovered  for the first time by infrared (IR) spectroscopy study of carbynes  in  \cite{Yearchuck_PL}.  Especially significant was the observation of  AFESWR with linear $k$-dipersion law, where  $k$ is magnitude of wave vector $\vec{k}$, that is the general spin wave theory, \cite{Hulthen}, \cite{Anderson},  \cite{F_Keffer},  \cite{Keffer}, \cite{Ziman}, \cite{Nakamura},  was experimentally confirmed  for AFESWR-case too. It is in agreement with the theoretical results represented in \cite{Yearchuck_PL}, from which is followed, that the conclusions of the  antiferroelectric spin wave  theory and antiferromagnetic spin wave theory are qualitatively identical, in particular, the same linear $k$-dipersion law is taking place (at low $k$-values).   

Discovery of new types of superconducting materials
has accelerated in 21th century. Especially interestig was   the discovery of superconductivity coexisting with antiferromagnetic ordering  in the iron-based
layered pnictide compound LaFeAsO (that is,  in material with prevailed 2D-dimensional strucure). It was repoted in \cite{Kamihara}. 
Next, the superconductivity has been discovered in both  oxygen containing RFeAsO (R = La, Nd, Sm) compounds and in oxygen free $AFe_2As_2$ (A = Ba, Sr, Ca)  compounds. It is interesting, that the  superconductivity
occurs upon doping into the FeAs layers of either electrons or holes. Let us remark, that owing to  the highly two-dimensional structure the pnictides are  like to the cuprates. It gave rise to the viewpoint
 that the physics of the
pnictides is similar to the cuprates. However, there is at present the dominating  viewpoint   that Mott-transition physics does not play a significant role
for the iron pnictides, and there are strong indications, that antiferromagnetic ordering is determined by the formation of the 
 spin-density wave (SDW), that is quite another  type of antiferromagnetism in comparison with Heisenberg
antiferromagnetism of localized spins takes place.  

At the same time on the coexistence of the superconductivity with spin wave resonance has not been reported. It wiil be reported in given work on the observation of uncompensated antiferromagnetic spin wave resonance, that is, in fact, ferrimagnetic spin wave resonance in carbon NTs, which is coexisting with  the superconductivity for the first time.

On the experimental revealing of magnetic ordering at all in carbon structurally ordered systems was reported  for the first time during the IBMM-Conference in Knoxville, TN, USA \cite{Erchak_Knoxville} and on E-MRS Conference in Strasbourg, France \cite{Efimov}. Let us remark, that almost in the same time   was repored  on magnetic ordering in structurally non-ordered carbon materials in  the work \cite{Kawataba}, where ferromagnetic ordering in pyrolytic carbon, produced  by chemical vapour deposition
 method was found.  Let us also remark, that simultaneously, the reports \cite{Erchak_Knoxville}, \cite{Efimov} were the first reports on the formation by high energy ion beam modification (HEIBM) of diamond single crystals  structurally and magnetically ordered quasi-one-dimensional (quasi-1D) system along ion tracks, that is, on the formation of new carbon allotropic form, which was identified with nanotubes (NTs), incorporated in diamond matrix in direction, which is precisely coinciding with ion beam direction.  Given NTs possess by a number of very interesting physical properties \cite{Erchak},
 \cite{Ertchak_Stelmakh}, \cite{Ertchak_JAS}. When concern the only magnetic ordering, it was established, that, for instance, incorporated nanotubes, produced by neon HEIBM of diamond single crystal along  $\langle{100}\rangle$ crystallographic direction, possess by weak antiferromagnetic ordering \cite{Erchak}, \cite{Ertchak_Stelmakh}, \cite{Ertchak_JAS}. At the same time, copper HEIBM with implantation direction along  $\langle{111}\rangle$ crystal axis,  nickel HEIBM with implantation direction along $\langle{110}\rangle$  axis \cite{Erchak}, \cite{Ertchak_Stelmakh}, \cite{Ertchak_JAS} and  boron HEIBM of polycrystalline diamond films with implantation direction transversely  to film surface \cite{Erchak_JETP} lead to formation of NTs, incorporated in diamond matrix, which  possesss by ferromagnetic ordering. It was established directly by observation of ferromagnetic spin wave resonance  \cite{Erchak_JETP}, \cite{Ertchak_Stelmakh}, \cite{Ertchak_JAS}.  It was found, that magnetic ordering is inherent property for given carbon electronic system and it is not connected with magnetic impurities. Very recently \cite{Yerchuck_D_Dovlatova_A}, antiferroelectric ordering has been found in the same pure carbon allotropic form - quasi-1D carbon zigzag-shaped  nanotubes (CZSNTs), obtained by boron- and copper-HEIBM of diamond single crystals in $\langle{111}\rangle$-direction. The proof for antiferroelectric ordering was obtained directly  by means of the detection of the new optical phenomenon - antiferroelectric spin wave resonance (AFESWR), which was theoretically described and experinmentally confirmed for the first time in \cite{Yearchuck_PL}.   
Recently, the physical origin of the mechanism of the formation of ferromagnetic ordering in carbon nanotubes  produced by high energy ion beam modification of diamond single crystals in $\langle{110}\rangle$ and $\langle{111}\rangle$ directions has been established. It is determined by asymmetry of spin density distribution  in Su-Schrieffer-Heeger topological soliton lattice formed in 1D Fermi quantum liquid state of the only  $\pi$-electronic subsystem of given NTs \cite{Yerchuck_D_Stelmakh_V_Dovlatova_A_Yerchak_Y_Alexandrov_A}. It was experimentally  proved, that  $\sigma$-electronic subsystem does not give any contribution to the mechanism of the formation of ferromagnetic ordering in given  NTs.

Quite other picture was observed  in  carbon NTs, produced by high energy ion beam modification of diamond single crystals in $\langle{100}\rangle$ direction. 
The  strong uncompensated antiferromagnetic  ordering {magnetic strength characteristics of which are comparable with magnetic strength characteristics of magnetic  ordering in atomic systems with unfilled or partly filled inner atomic shells] coexisting with superconductivity at room temperature  is argued in the work \cite{Yerchuck_Stelmakh_Dovlatova_Yerchak_Alexandrov}.
  The mechanisms of superconducting state formation are proposed to be the following. On the one hand,  s-wave mechanism, mediated by the coupling of charge carriers with stretched phonon modes like  to those ones, taking place in $MgB_2$ \cite{Nagamitsu}, heavily boron doped diamond \cite{Ekimov}, \cite{Takano},
 \cite{Blase}, \cite{Lee} and  sandwich S-Si-QW-S structures \cite{Bagraev} seems to be realised. Moreover, just crimped cylindrical shape allows to increase the strength of C-C bonds by preservation of high density of the states on Fermi surface, resulting from low dimensionality (quasi-1D) of NTs. On the 
`other hand, the multiband structure of valence and conductivity bands allows to realise the formation of joint antiferromagnetic-superconducting state by means of the $s{\pm}$-wave  and $p$-wave formation like to pnictides.  Taylor expansion over dimerization coordinate of electron-electron interaction and electron-phonon interaction indicate on the possibility of the  formation of superconducting state by different channels. The independent on dimerization coordinate (which can be both in static and dynamic states)  electron-electron repulsion terms can give the contribution to antiferromagnetic-superconducting $s^+$-state formation like to pnictides. The attractive terms, which are proportional to dimerization coordinate, can lead to formation of superconducting s-state by two   mechanisms. The first mechanism is  like to BCS-superconductivity mechanism \cite{BCSch}. The second mechanism is new. It is mediated by the coupling of charge carriers with stretched phonon modes in C-C bonds. 

Especially interesting seems to be the role of external quantized EM-field, since the only in resonance conditions the switch to superconducting state was  realised.  It was explained by appearance by resonance  of long-lived  coherent system of resonance hypersound phonons. It means, that quantized radiospectrospy-range EM-field has to be working constituent for realisation of oom temperature superconducting state. It was suggested, that the room temperature superconducting state in $\langle{100}\rangle$-NTs, incorporated in diamond matrix  is the  superposition of a number of superconducting-wave  states above indicated.

The main aim of the work presented is to obtain experimentally the  exact experimental proof 
for the conclusions in \cite{Yerchuck_Stelmakh_Dovlatova_Yerchak_Alexandrov}. 

\section{Experimental Technique}

The same samples, that in \cite{Yerchuck_Stelmakh_Dovlatova_Yerchak_Alexandrov}, that is the samples of type IIa natural diamond, implanted by high energy ions of nickel (the energy of ions in ion beam was  $335$ $MeV$) have been studied. The absolute spin number in each of the samples studied did not exceed before implantation the value $\approx 10^{12}$ spins. Therefore, the samples were magnetically pure samples.  Ion implantation (ion beam dose was $5\times{10^{13}}$ $cm^{-2}$) was performed along $\left\langle{100}\right\rangle$ crystal direction, that is,  transversely to sample (100)-plane  uniformly along all the  plane surface. The temperature of the samples during the implantation was controlled and it did not exceed 400 K.

  X-band ESR-spectrometer "Radiopan" was used for the registration of  magnetic resonance spectra. They were registered  by using of $TE_{102}$ mode rectangular cavity at room temperature. The ruby standard sample was permanently placed in the cavity on its sidewall. One of the  lines of ESR absorption by  $Cr^{3+}$ point paramagnetic centers (PC) in ruby  crystal was used  for the correct relative intensity measurements of resonance absorption, for the calibration  of the amplitude value of magnetic component of the microwave field and for precise phase tuning of modulation field. The correct relative intensity measurements become to be  possible owing to unsaturating behavior of ESR absorption in ruby in the range of  the microwave power applied, which was  $\approx  100$ mW in the absence of attenuation. Unsaturable character of the absorption in a ruby standard was confirmed by means of the measurements of the absorption intensities in two identical ruby samples in dependence on the microwave power level. The first sample was standard sample,  permanently placed in the cavity, the second sample was placed in the cavity away from the loop of magnetic component of microwave field  so, that  its resonance line intensity was about 0.1 
of the intensity of corresponding line of the standard sample. The slightly different  orientation of the samples allowed the simultaneous registration of both the samples without overlapping of  their absorption lines.  The  foregoing intensity ratio about 0.1 was precisely preserved for all microwave power values in the range used. Consequently, ruby samples are realy good standard samples by ESR spectroscopy studies.

 \section{Results}
The ESR spectrum observed in carbon nanotubes, produced by nickel  high energy $\left\langle{100}\right\rangle$ ion beam modification  of natural diamond single crystals, is presented in Figure 1  in  crystal direction [111]. It was reported in \cite{Yerchuck_Stelmakh_Dovlatova_Yerchak_Alexandrov}, that the  spectrum observed  was excited spontaneously the only by very precise orientation of external static magnetic field $\vec{H}_0$ in the 0$\overline{1}$1) plane  of the sample and that resulting spectrum was consisting of three clearly pronounced lines, at that two from given new lines have rather large anisotropic linewidths and from two very broad lines, which were registered only partly in the range of magnetic fields 0-4000 G. We will preserve the designations for three clearly pronounced lines, which were used in \cite{Yerchuck_Stelmakh_Dovlatova_Yerchak_Alexandrov}, that is $R_b$  for the relatively right broad line and by L for the more broad left line.  The right broad line was overlapped with relatively narrow almost isotropic line, designated by $R_n$ (given line was observed by usual(that is not very precise) sample orientation). Very broad strongly intensive anisotropic absorption can be characterised by  two dip positions (in integrated spectrum) at $\sim{2410}$ G and $\sim{2892}$ G (corresponding to to different lines by spectrum registration in the direction coinciding with [111] diamond lattice direction, Figure 1. It was established in \cite{Yerchuck_Stelmakh_Dovlatova_Yerchak_Alexandrov}, that two dip positions for given background absorption were coinciding by static magnetic field   direction in 60 degrees from [100] diamond crystal direction, that was argued to be the  display of the fact, that the symmetry of the interaction, leading to the appearance of very strong  background absorption is determined by inherent magnetic symmetry of NTs, produced by [100] HEIBM, which is not coinciding with structure symmetry  of given NTs. 

Let us summarise to better understanding of subsequent discussion the  experimental result,described in \cite{Yerchuck_Stelmakh_Dovlatova_Yerchak_Alexandrov}. Dependencies of absorption amplitudes of L-line and  $R_b$ line on magnetic component of microwave field  at fixed orientation of polarising magnetic field $\vec{H_0} || ${[100]} crystal axis have been studied. Given dependencies were quite different for L-line and  $R_b$ line. The dependence for L-line is superlinear. It is similar to the dependencies, which were earlier observed in the samples, modified by HEIBM with copper, neon, nickel ions (however with dose $5 \times 10^{14}$) \cite{Erchak}, \cite{Ertchak_Stelmakh}, \cite{Ertchak_JAS}, that is, in the case of entire modification of diamond layer, which is localised near surface. In that case the layer consists the only of NTs, which seem to be  interacting each other, that in its turn leads to more short spin-lattice  relaxation time $T_1$ for individual spin carrier (see for more detailed explanation the papers \cite{Erchak}, \cite{Ertchak_Stelmakh}, \cite{Ertchak_JAS}.  In the studied sample (integral  dose is $5 \times 10^{13}$), individual NTs are isolated by diamond structure, nevertheless the superlinear dependence is taking place, indicating on another mechanism of $T_1$  shortening.  Dependence of absorption amplitude of the right broad  line $R_b$ in ESR spectrum of NTs on magnetic component of microwave field was found to be  strongly nonlinear. It is characterised for the values of relative  magnetic component of microwave field $H_1/H_1^{(0)}$
in the range (0-0.75) by usual saturating law, but in  the range (0.75-1) it acquires prominent superlinear nonsaturating character \cite{Yerchuck_Stelmakh_Dovlatova_Yerchak_Alexandrov}. Given dependence was observed in ESR-spectroscopy for the first time.  Angular dependence of g-factor of the   line L consists of two branches. One branch is  in the angle range 0-60  degrees from [100] crystal lattice direction (which is coinciding with NT axis direction), the second branch is in the angle range 60-90 degrees. 
It was remarked in \cite{Yerchuck_Stelmakh_Dovlatova_Yerchak_Alexandrov}, that the connection point of two branches, equaled to 60 degrees for the g-values of L-line is coinciding with the point of the junction of two dips in the very broad (and consequently very intensive) absorption, testifying on the same (or related) nature of the resonance processes, which are responsible for the appearance of L-line and very broad lines.  It was also found, that the deviation of g-values from free electron value g = 2.0023 is very large, at that the minimal value is achieved in the range 16-20  degrees from the [011]  direction in diamond lattice and it is equal to $\approx 2.0719$, maximal g-value corresponds to NT axis direction, that is to [100] crystal lattice direction and it is  
equal to $\approx 2.3120$. Given values are characteristic for the systems with the strong magnetic ordering. Given data were interpreted to be the direct proof of the spontaneous transition of NT system, incorporated in diamond lattice, in the state with the strong magnetic ordering.   Angular dependence of ESR absorption intensity of the   line L has qualitatively opposite character to g-factor dependence. The maximal absorption value corresponds to the direction, being to be transversal to NT axis, which is coincides with  [011]  direction in diamond lattice.  Additional maximum is observed at 60 degrees from given direction. It is characteristic, that both the maxima in angular dependence of  absorption intensity of the   line L are observed also in angular dependence of its linewidth. It is indication, that the main features in angular dependence of ESR absorption intensity are governed by angular dependence of linewidth. Especially interesting, that the line L is asymmetric with values of the ratio A/B of the asymmetry extent, which are  quite different in comparison with those ones by usual Dyson shape \cite{Dyson}. Let us remark, that if Dyson effect is taking place, then by usual tuning of microwave phase on absorption  registration the value A/B is equal to 2.55 for derivative of the responce signal in the case of  static (immobile) paramagnetic centers (PC) in conductive media when  the samples are  thick in comparison with the scin depth. It is determined by the space dispersion contribution \cite{Erchak_Zaitsev_Stel'makh}, which is appeared by ESR detection in conductive media. It corresponds to the ratio of space dispersion contribution and absorption contribution to resulting ESR response equaled to (1 : 1) \cite{Erchak_Zaitsev_Stel'makh}.  The value A/B for absorption derivative is increasing from 2.55 to more than 19 for mobile PC in dependence on velocity of the spin diffusion \cite{Poole}. In the case of thin samples  the ratio A/B has intermediate values, between 1 and above indicated, depending on the thickness of the samples. It was found, that the ratio A/B is less than 1 and it is strongly anisotropic. The maximal A/B  value is near [111] crystal lattice  direction and it is equal to 0.83. The minimal A/B  value is near 60 degrees from [011] crystal lattice  direction and it is equal to 0.49. It was also remarked, that by  Dyson effect in conducting samples (in particular in the samples with metallic NTs, producing the network) the maximal deviation from the ratio A/B = 1 has to be observed by  microwave field propagation direction along the sample side with maximal size, that is by $H_0$ along [100] crystal direction,  in the case, when the network is opaque for  microwave field in direction, transverse to NT axis direction, or  by $H_0$ along [011] in the case, when the network is opaque  the only for  microwave field propagation in direction, coinciding with NT axes. The observed maximal deviation of  ratio A/B  from A/B = 1 at $\approx{60}$ degrees from [011] confirms the conclusion on nontrivial nature of Dyson-like effect in the case studied. 

  It has been found to be substantial, that Q-factor is increasing in the ranges,  where deviation of  ratio A/B  from A/B = 1 is also increasing, that is, increase is starting near 60 degrees from [011] crystal lattice  direction and increase takes place in the range near 10-30 degrees from the same [011] crystal direction. For usual Dyson effect it has to be conversely, Q-factor has to be minimal in the direction of maximal  deviation of  ratio A/B  from A/B = 1, that is near 30  degrees from [100] crystal lattice  direction. However, it was found, that Q-factor has in given direction the maximal value. 

It was argued in \cite{Yerchuck_Stelmakh_Dovlatova_Yerchak_Alexandrov}, that the results above described are agreed with spontaneous transition of the system to the state which characterises by coexistence simultaneously of antiferromagnetic (AFM) ordering and superconductivity, which is realized in electron spin resonance conditions and it is absent without resonance. It was also concluded, that the 
  nature of given state and mechanisms, leading to its formation cannot be entirely coinciding with all known ones.
Really, the suggestion  on just  AFM ordering (but not ferromagnetic) is in agreement with observation of two both very broad and moderately broad lines. 
The appearance of two resonance lines (if linearly polarised microwave field is used by detection) was established by Kittel in the work \cite{Kittel}, which was the first work on the theory of AFM-resonance. We  have found, that magnetic moments of two sublattices being to be opposite directed are uncompensated in their magnitude, that is, strongly speaking, we are dealing with 
uncompensated AFM-resonance or in other words with ferrimagnetic resonance. This is so indeed, since the ratio of intensities of the absorption, corresponding to 
L and $R_b$-lines is equal to $\approx 3.5$.

\begin{figure}
\includegraphics[width=0.5\textwidth]{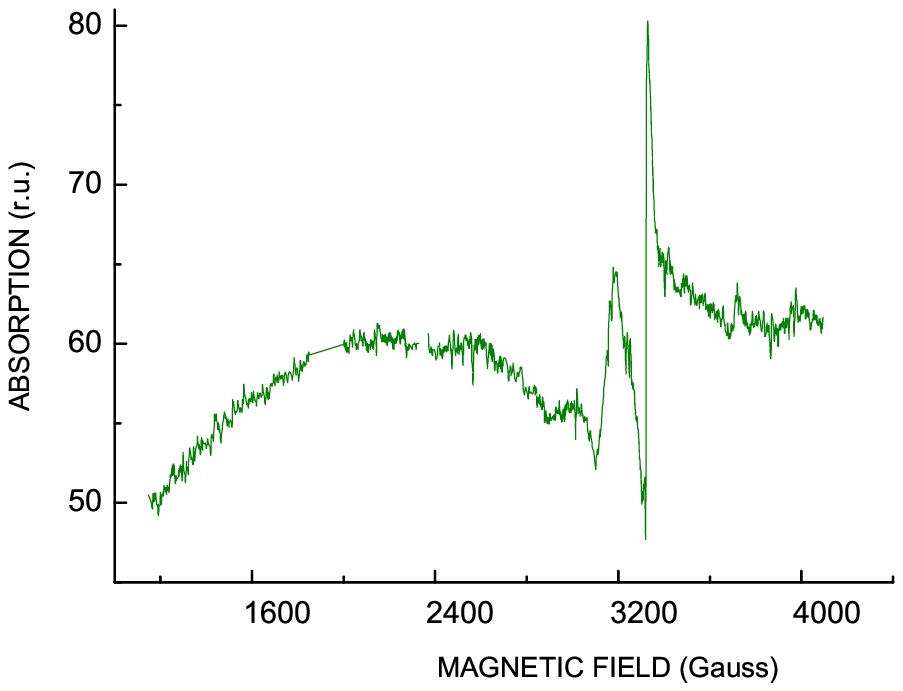}
\caption[Spectral distribution of ESR absorption  intensity in
diamond single crystal, implanted by high energy nickel ions  by   beam direction transversely (100) sample plane, the sample was rotated in (0$\overline{1}$1) plane, $\vec{H_0} ||$ {[111]} crystal axis]
{\label{Figure1} Spectral distribution of ESR absorption  intensity in
diamond single crystal, implanted by high energy nickel ions  by   beam direction transversely (100) sample plane, the sample was rotated in (0$\overline{1}$1) plane, $\vec{H_0} ||$ {[111]} crystal axis}\end{figure}
\begin{figure}
\includegraphics[width=0.5\textwidth]{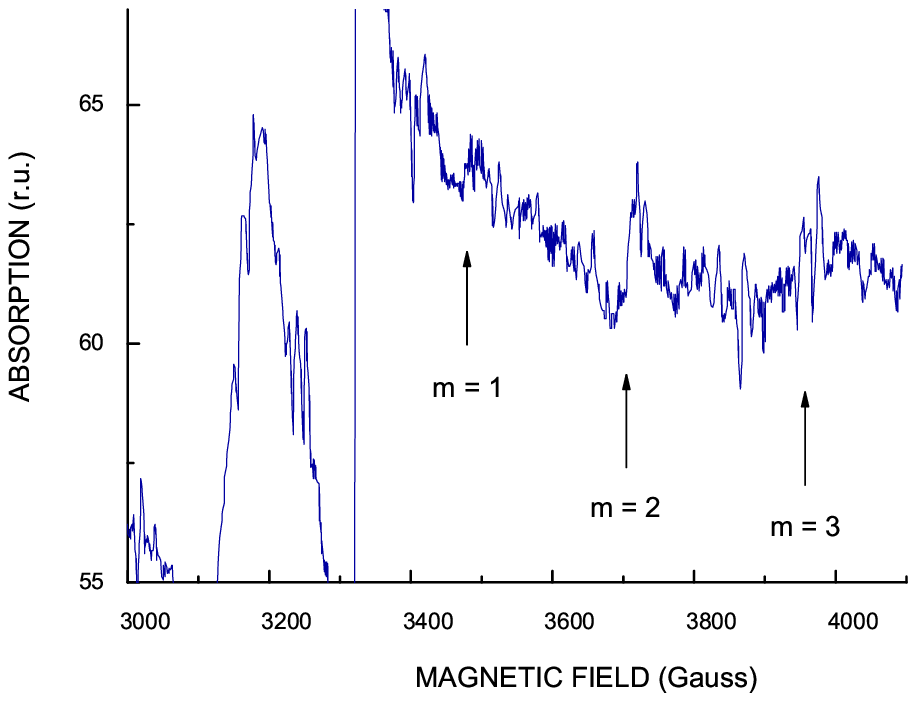}
\caption[Detailed spectral distribution of  magnetic resonance absorption  intensity in the range 3000 - 4100 Gauss in 
diamond single crystal, implanted by high energy nickel ions  by   beam direction transversely (100) sample plane, the sample was rotated in (0$\overline{1}$1) plane, $\vec{H_0} ||$ {[111]} crystal axis]
{\label{Figure2} Detailed spectral distribution of  magnetic resonance absorption  intensity in the range 3000 - 4100 Gauss in 
diamond single crystal, implanted by high energy nickel ions  by   beam direction transversely (100) sample plane, the sample was rotated in (0$\overline{1}$1) plane, $\vec{H_0} ||$ {[111]} crystal axis}\end{figure}

\begin{figure}
\includegraphics[width=0.5\textwidth]{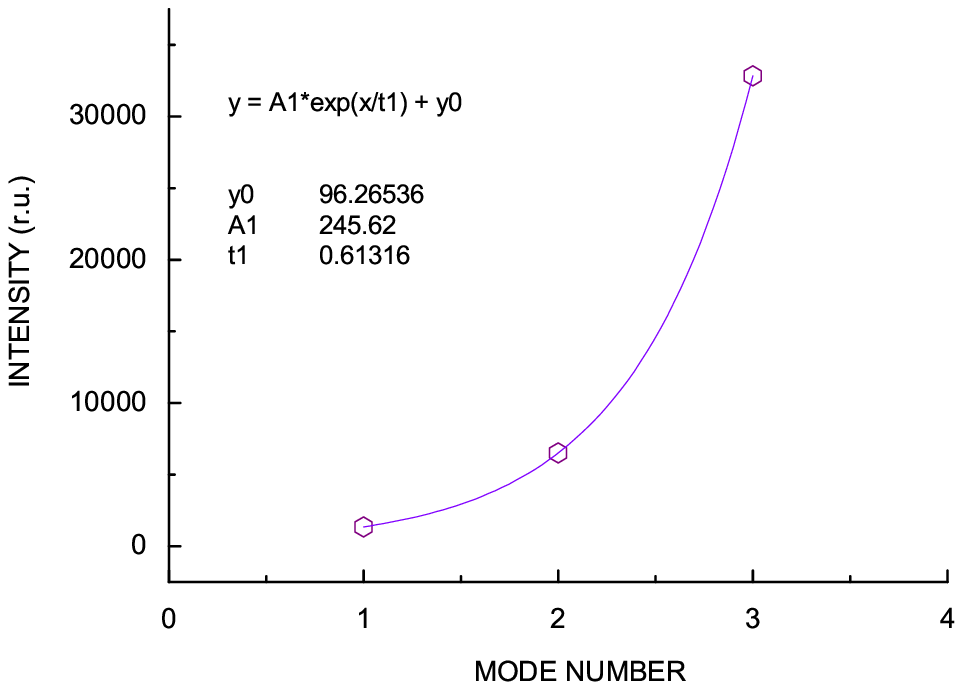}
\caption[Dependence of AFMSWR mode intensity on mode number in
diamond single crystal, implanted by high energy nickel ions  by   beam direction transversely (100) sample plane, the sample was rotated in (0$\overline{1}$1) plane, $\vec{H_0} ||$ {[111]} crystal axis]
{\label{Figure3}  Dependence of AFMSWR mode intensity on mode number in
diamond single crystal, implanted by high energy nickel ions  by   beam direction transversely (100) sample plane, the sample was rotated in (0$\overline{1}$1) plane, $\vec{H_0} ||$ {[111]} crystal axis}\end{figure}

The experimental spectra were analysed more carefully. It has been found, that along with main uncompensated AFM-modes the uncompensated AFM spin wave resonance (SWR) modes are also excited. They are clearly registered on right side from main uncompensated AFM-modes, see Figure 2, where the modes with the numbers $m = \overline{1,3}$ are represented. The splitting between the modes is the following:   
between the center of "gravity" of two main AFM-modes and the first SWR-mode the splitting is equal to $246 \pm\ 5 G$, between the first SWR-mode and the second  SWR-mode it is  equal to $228 \pm\ 10 G$ and the distance between the  second and the third SWR-modes is  equal to $251 \pm\ 15 G$. Therefore, we have nearly equidistant SWR splitting, that is SWR with practicaly linear dispersion law and consequently the direct proof for the formation of uncompensated  AFM ordering.

 The shape of SWR-modes and their intensity distribution were analysed. Taking into account the experimental distribution for the values of amplitudes $A_i$, $i = \overline{1,3}$, of SWR-modes $A_1$:$A_2$:$A_3$ = 1.06 : 2.65 : 1.83 and the linewidth $\Delta H_j$, $j = \overline{1,3}$, $\Delta H_1 = 35.7 \pm 3 G$, $\Delta H_2 = 49.6 \pm 5 G$, $\Delta H_3 = 134 \pm 8 G$, we obtain the intensity distribution, represented in Figure 3. It is seen, that the strong  growth of intensity is observed with the mode number increasing. So, the intensity of the third mode is greater by a factor of 24.3  of the  intensity of the first mode. Given growth can be approximated by exponential dependence, although it is evident, that  the number of experimental points is insufficient upto insist, that given law is really takes place. Nevertheless, the strong  growth of intensity with mode number   is unusual for all earlier known magnetic and electrical SWR resonances, for which the intensity of the mode with greater number is not exceeding the  intensity of the mode with lower number, see, for instance, \cite{Seavey_Tannenwald}, \cite{Yearchuck_Yerchak_Doklady}, \cite{Yearchuck_PL}. It seems to be very useful to analyse the  asymmetry extent A/B  of the modes observed. It was done for the only second and the third modes. It is interesting, that the deviation of asymmetry extent from 1 is positive in distinction from the case of the main AFM-modes. The ratio A/B is equal $1.07 \pm 0.03$ for the second mode and it is equal $1.25 \pm 0.05$ for the third mode. Therefore, there is clear tendency to the increase of asymmetry extent with mode number increase.

\section{Discussion}

It will be further argued, that the additional results above described confirm the preliminary conclusions of the work \cite{Yerchuck_Stelmakh_Dovlatova_Yerchak_Alexandrov}, that is spontaneous transition of the system studied to the state which characterises by coexistence simultaneously of uncompensated AFM ordering and superconductivity is really takes place in magnetic resonance conditions.
We will show, that the peculiarities of angular dependences of cavity Q-factor and A/B ratio, the numerical  values of A/B ratio for both main uncompensated AFM-modes and AFMSWR modes are not connected with usual Dyson effect and they are determined by quite other mechanism.

 It is known, that the dynamical spin susceptibility of a superconductor with $s^+$ cannel is
given by an  formula, obtained by random phase approximation  (RPA) method  in 
\cite{Korshunov_Eremin}. Within RPA the spin response has an operator form. It is

\begin{equation}
\label{eq1}
\hat{\chi}_s(\vec{q}, \Omega) = [\hat{I} - \hat{\Gamma} \hat{\chi}_s^{0}(\vec{q}, \Omega)]^{-1}\hat{\chi}_s^{0}(\vec{q}, \Omega), 
\end{equation}
where $\hat{I}$ is  the unit operator, $\hat{\Gamma}$ is  the coupling  operator, $\hat{\chi}_s^{0}(\vec{q}, \Omega)$
 is the operator, formed by the interband and intraband bare susceptibilities. Given relation was analysed in the work \cite{Chubukov} in application to iron based superconductors. However, the result obtained will be qualitatively true for any superconductor with $s^+$ cannel.
It has been found, that in the normal state, the observable quantity, corresponding to dynamical spin susceptibility operator  is
 only weakly logarithmically depends on frequency.
 In a superconducting state, it has a resonance behavior.
In the absence of SDW instability the authors of \cite{Chubukov} have obtained the following expression for resonance frequency
\begin{equation}
\label{eq2}
\Omega = \sqrt{v^2(\vec{q} - \vec{Q})^2 - (\Omega_0)^2},
\end{equation}
where $v = \frac{v_F}{\sqrt{2}}$, $v_F$ is Fermi velocity, $\vec{Q}$ is momentum, corresponding to point in  the folded Brillouin
zone, around which electron pockets are 
centered. 
$\Omega_0$  in (\ref{eq2}) is
\begin{equation}
\label{eq3}
\Omega_0 = 2\Delta \sqrt{(\Gamma^r_{SDW})^{-1} - \log{\frac{E_F}{E_0}}},
\end{equation}
where $E_F$ is Fermi energy, $\Gamma^r_{SDW}$ is factor, characterising interband interaction and $E_0$ is the largest of superconducting gap $\Delta$
and the cutoff energy associated with nonequivalence of
the Fermi surfaces for electrons and holes.  The shape of given resonance was represented in Figure 3 of \cite{Korshunov_Eremin}, It is seen, that the shape is strongly asymmetric. The authors have also found that the resonance peak is confined
to the AFM wave vector $\vec{Q}$ and disappears rapidly for
$|\vec{q}| < |\vec{Q}|$. So  already at $|\vec{q}| = 0.995 |\vec{Q}|$ the susceptibility  is much smaller than its value at $|\vec{Q}|$. Physically the appearance of magnetic resonance behavior in $s^+$ superconductors is determined by the presence of the magnetic fluctuation spectrum
consiststing of  the continuum of
the AFM spin fluctuations peaked at $\vec{Q}$ and which
 arise  to be the consequence of the interband scattering.  

Spin resonance in $s^+$ superconductor was compared in \cite{Korshunov_Eremin} and in \cite{Chubukov} with the spin resonance in  $d_{x^2-y^2}$ superconductors.  Both the resonances have the  similarities
and differences.
On  the one hand, they are excitonic resonances, and they occur
because the superconducting  gap changes sign between the Fermi surface points with momenta $\vec{k}$ and
$\vec{k} + \vec{Q}$. On the other hand, the resonance frequency in a $d_{x^2-y^2}$ superconductors
disperses downward because of the nodes, while for a
nodeless $s^+$ superconductor, the resonance disperses upward, with large
velocity. Concerning our data, it is seen from Figure 1 in given paper and from Figures 1 to 3 and Figure 9 in \cite{Yerchuck_Stelmakh_Dovlatova_Yerchak_Alexandrov}, that symmetry extent of resonance line L is in agreement with the formation of $s^+$ superconducting state in the sample studied. Really, to   more intensive right side of resonance line by its registration with the frequency scan will correspond the line with more intensive left side by its registration with the fixed frequency and by scan of static magnetic field.

Therefore, asymmetry of spin resonance in superconductors has quite other origin in comparison with Dyson effect in metals (or other conducting media). We will show that AFMSWR resonance in $s^+$ superconductors has  also peculiarities, which allow to detect the superconducting state. They are the following.

1.The change of asymmetry extent of resonance modes in comparison with main AFM  modes (positive deviation from A/B = 1) and its increase with mode number increase. 

It can be explained by the fact, that AFMSWR modes have nodes, that, like to  the resonance lines in a $d_{x^2-y^2}$ superconductors become the opposite asymmetry in comparison with $s^+$ nodeless wave registered directly by AFM resonance. It is clear, that positive deviation from 1 will growth with increse of node number, that is, with mode number increase, that actually takes place.
2.The substantial growth of the intensity of the AFMSWR modes with mode number increase, see Figure 3. 

Let us remark that intensity conservation law for SWR modes was found  for NTs incorporated in diamond matrix with other implantation directions \cite{Ertchak_JAS}, carbynes and for some organic quasi-1D substances (polyvinylidenehalogenides - PVDF) \cite{Yearchuck_PL}. At that, there are similarities in spectroscopic
properties between FMSWR, observed by ESR in ferromagnetically
ordered 1D-lattices of topological solitons
(SSH-solitons or those ones, belonging to SSH-class) and antiferroelectric SWR
(AFESWR). So, amplitude of FMSWR modes are decreasing
with mode number  $m$ proportionally $\frac{1}{m}$. At the same time, the linewidths
are increasing by such a way, that intensity of the modes
is practically conserved \cite{Ertchak_JAS}. It holds by AFESWR also
true. So, in \cite{Yearchuck_PL} is given the concrete example - the ratio of relative amplitudes of the first to
the second AFESWR modes  in PVDF sample, is $1.9
(± 0.2$) $cm^{-1}$. The linewidths are $19$ and $37 (± 1.5)$  $cm^{-1}$, which
gives the same ratio for the linewidths of the second to
the first modes.  Given property was considered  to be sufficient
 to insist, that
topological quasiparticles are responsible for SWR. In the other earlier known cases, for instance by FMSWR in ferromagnetic metals, the intensity of SWR modes is decreasing with mode number increasing, see, for instance, Figure 1 in \cite{Seavey_Tannenwald}.

 The substantial increase of the intensity of AFMSWR modes with mode number increasing becomes to be understandable, if to take into account the presence of the magnetic fluctuation spectrum consisting of  the continuum of
the AFM spin fluctuations peaked at $\vec{Q}$. For AFMSWR modes $|\vec{q}| \neq 0$ and $|\vec{q}|$ is increasing with mode number increasing coming  near to the value of $\vec{Q}$. Then the dynamical magnetization will be determined by Fourier component of the magnetic fluctuation field on the operating microwave frequency of the spectrometers, which is added to dynamical magnetization produced by magnetic component of microwave field used.  

Therefore, we obtain the direct experimental proof of the formation of superconducting $s^+$ cannel in the sample studied. The possibility of the realization of another cannels and the role of resonance conditions are discussed in  \cite{Yerchuck_Stelmakh_Dovlatova_Yerchak_Alexandrov}.

\section{Conclusions} 

The phenomenon  of ferrimagnetic  spin wave resonance [uncompensated antiferromagnetic spin wave resonance] has been detected for the first time. 
Given phenomenon was observed in carbon nanotubes, produced by high energy ion beam modification of diamond single crystals in $\langle{100}\rangle$ direction. The fact itself of observation of  uncompensated antiferromagnetic spin wave resonance is direct proof of the formation of antiferromagnetic ordering [uncompensated], which is found rather strong. It is comparable with magnetic ordering in classical magnetic substances, elementary units of which contains the elements with unfilled inner atomic shells. Given property of carbon is established  also for the first time. Spin wave resonance observed has two main peculiarities.  

1.The opposite deviation of the asymmetry extent ratio A/B from 1 of  resonance modes in comparison with main AFM mode, at that it increases with mode number increase. It is explained qualitatively by existence of nodes like to explanation of the asymmetry extent of  the resonance lines in a $d_{x^2-y^2}$ superconductors. 

2.The substantial increase of the intensity of AFMSWR modes with mode number increase. It is explained by  taking into account the presence of the magnetic fluctuation spectrum consisting of  the continuum of
the AFM spin fluctuations peaked at AFM vector $\vec{Q}$. For AFMSWR modes wave vector $|\vec{q}| \neq 0$ and $|\vec{q}|$ is increasing with mode number increase, coming  near to the value of $\vec{Q}$. Then the dynamical magnetization will be determined by Fourier component of the magnetic fluctuation field with the frequency, coinciding with the  operating microwave frequency of the spectrometer. Given component is added to dynamical magnetization produced by magnetic component of microwave field used and it determines mode intensity growth.

The peculiarities of  AFMSWR above indicated allow to insist on the formation in given nanotubes of   $s^+$-superconductivity at room temperature, coexisting with uncompensated antiferromagnetic ordering.

\end{document}